\documentclass[12pt,a4paper]{article}
\usepackage{amsmath}
\usepackage{amssymb}
\usepackage{latexsym}
\begin{document}

\title{Symmetry of $osp(m|n)$ spin Calogero-Sutherland models}
\author{Kazuyuki Oshima\\
 \\
 Aichi Institute of Technology\\
 1247 Yachigusa, Yakusa Cho, Toyota City,\\
 Aichi Prefecture 470-0392, Japan \\
 e-mail: oshima@aitech.ac.jp}

\date{\empty}

\maketitle
 
\begin{abstract}
 We introduce $osp(m|n)$ spin Calogero-Sutherland models and find that
the models have the symmetry of $osp(m|n)$ half-loop algebra or
Yangian of $osp(m|n)$ if and only if the coupling constant
of the model equals to $\frac{2}{m-n-4}$.
 
\bigskip


\bigskip


\end{abstract}

\section{Introduction}
The Calogero-Sutherland models are one-dimensional many particle systems with
long range interactions. We denote by $L$ and $\lambda$ the number of 
particles and the coupling constant which determines the strength of 
the interaction, respectively. The Hamiltonian of the model is expressed as
\begin{equation}
 H=- \sum_{j=1}^{L} \frac{\partial^2}{\partial x_{j}^{2}}
   + 2 \lambda \sum_{j < k} (\lambda-1) V(x_{j}-x_{k})
\end{equation}
where the potential $V(r)$ is $1/r^2$ (rational), 
$1/\sin^2 r$ (trigonometric), and $\wp (r)$ (elliptic).
We often call the rational case and the trigonometric case 
the Calogero model and the Sutherland model respectively.
There are various generalizations to the Calogero-Sutherland models. 
One of the generalizaitons is the spin generalization, 
namely, we consider models for which 
particles have $gl(N)$ spin as an internal degree of freedom. 
The Hamiltonian is
\begin{equation}
 H=-\sum_{j=1}^{L} \frac{\partial^2}{\partial x_{j}^{2}}
   + 2 \lambda \sum_{j < k} (\lambda-P_{jk}) V(x_{j}-x_{k}),
\end{equation}
where $P_{jk}$ is a permutation operator in a spin space, and exchange
the spin state of the $j$-th particle and the $k$-th particle.
Using the spin operator $e^{ab}$ as a basis of $gl(N)$, the operator
$P_{jk}$ can be written as
\begin{equation}
 P_{jk}=\sum_{a,b=1}^{N} e_{j}^{ab} \otimes e_{k}^{ba}.
\end{equation}
The symmetries of the models turn to be the half-loop algebra 
or the Yangian of $gl(N)$ \cite{BGHP}\cite{HW1}\cite{HW2}\cite{BHW}.
This $gl(N)$ spin Calogero-Sutherland models have supersymmetric 
extensions, which are what we call $gl(m|n)$ spin Calogero-Sutherland 
models \cite{AK}\cite{JCGWW}\cite{JWW}. It is also proved that the $gl(m|n)$ spin 
Calogero-Sutherland models have the Yangian $Y(gl(m|n))$ symmetry.
Recently new interactions between the internal degree
of freedom were introduced in \cite{C}. These interaction are defined 
in terms of the fundamental representaiton of
the generators of Lie algebra $so(N)$ or $sp(N)$. Then we call these models $so(N)$ or
$sp(N)$ spin Calogero-Sutherland models. It is shown that 
the $so(N)$ or $sp(N)$ spin Calogero-Sutherland models have symmetry algebras
if and only if the coupling constant takes a particular value. 

It is natural to ask if the $so(N)$ or $sp(N)$ spin Calogero-Sutherland models
have supersymmetric extensions.
The purpose of this paper is to extend the $so(N)$ or $sp(N)$ spin Calogero-Sutherland 
models to the Lie superalgebra $osp(m|n)$ case, namely the particles carry
the internal degree of freedom which is described in terms of a
representation of the orthosymplectic Lie superalgebra $osp(m|n)$.
We show that our models have the half-loop algebra of $osp(m|m)$ or the Yangian of 
$osp(m|n)$ as the symmetry algebra when the coupling constant equals to 
$\frac{2}{m-n-4}$.

This paper is organized as follows. In section 2, we define the orthosymplectic
Lie superalgebra $osp(m|n)$. Then we introduce a new model called $osp(m|n)$
spin Calogero model in section 3. We find the symmetry of the $osp(m|n)$
spin Calogero models in section 4. In section 5, we consider the trigonometric
case, that is, $osp(m|n)$ spin Sutherland models. Finally we show 
that the $osp(m|n)$ spin Sutherland models have super Yangian $Y(osp(m|n))$ symmetry.

\section{Orthosymplectic Lie superalgebra}
In this section we will give the fundamental notations of the Lie superalgebras.
For details, see \cite{GZ}, \cite{W} for example.
Throughout this paper, we assume $n$ is even.
Let $e^{ab}$ be the standard generators of $gl(m|n)$, the 
$(m+n) \times (m+n)$-dimensional general linear Lie superalgebra, 
obeying the graded commutation relations
\begin{equation}
 \left[ e^{ab}, e^{cd} \right] = 
 \delta_{bc} e^{ad} -(-1)^{([a]+[b])([c]+[d])} \delta_{da} e^{cb}
\end{equation}
where $[a]$ is the $\mathbb{Z}_{2}$ grading defined as
\begin{displaymath}
 [a] = \left\{
         \begin{array}{ll}
           0, & a=1, \dots ,m \\
           1, & a=m+1, \dots , m+n.
        \end{array}
        \right. 
\end{displaymath}
The orthosymplectic Lie superalgebra $osp(m|n)$ is a subsuperalgebra
of the general linear Lie superalgebra $gl(m|n)$. 
Using the generators $e^{ab}$ of $gl(m|n)$, we can construct $osp(m|n)$ as follows.
For any $a=1, \dots, m+n$, we introduce a sign $\xi_{a}$
\begin{displaymath}
  \xi_{a} = \left\{
         \begin{array}{ll}
           +1, & 1 \le a \le m+ \frac{n}{2} \\
           -1, & m+\frac{n}{2}+1 \le a \le m+n
        \end{array}
        \right. 
\end{displaymath}
and a conjugate ${\bar a}$
\begin{displaymath}
 {\bar a} = \left\{
         \begin{array}{ll}
           m+1-a, & a=1, \dots, m \\
           2m+n+1-a, & a=m+1, \dots , m+n.
        \end{array}
        \right. 
\end{displaymath}
Note that
\begin{equation}
 \xi_{a}^2=1, \quad \xi_{a}\xi_{{\bar a}}=(-1)^{[a]}.
\end{equation}
Then we choose an even non-degenerate supersymmetric metric $g_{ab}$
as follows,
\begin{equation}
 g_{ab}= \xi_{a} \delta_{a {\bar b}},
\end{equation}
with inverse metric 
\begin{equation}
 g^{ba}=\xi_{b} \delta_{b {\bar a}}.
\end{equation}
As generators of the orthosymplectic Lie superalgebra $osp(m|n)$ we take
\begin{equation}
 \sigma^{ab}=g_{ak}e^{kb}-(-1)^{[a][b]}g_{bk}e^{ka} =-(-1)^{[a][b]}\sigma^{ba},
\end{equation}
which satisfy the graded commutation relations
\begin{eqnarray}
 [ \sigma^{ab}, \sigma^{cd} ] &=& g_{cb} \sigma^{ad}
  -(-1)^{([a]+[b])([c]+[d])}g_{ad}\sigma^{cb} \nonumber \\
  && -(-1)^{[c][d]}(g_{db} \sigma^{ac} 
   -(-1)^{([a]+[b])([c]+[d])}g_{ac}\sigma^{db} ).
\end{eqnarray}
It is easy to check that these generators satisfy the following equations:
\begin{eqnarray}
 \left[ \sigma^{ab}, \sigma^{cd} \right] &=&
  -(-1)^{([a]+[b])([c]+[d])} \left[ \sigma^{cd}, \sigma^{ab} \right], \\
 \left[ \left[ \sigma^{ab}, \sigma^{cd} \right], \sigma^{ef} \right] &=&
 \left[ \sigma^{ab},\left[ \sigma^{cd}, \sigma^{ef} \right] \right] \nonumber \\
  && \quad -(-1)^{([a]+[b])([c]+[d])} 
   \left[ \sigma^{cd},\left[ \sigma^{ab}, \sigma^{ef} \right] \right].
\label{superJ}
\end{eqnarray}
These relations are the defining relations of the Lie superalgebras.
The relation (\ref{superJ}) is called the super Jacobi identity.

\section{$osp(m|n)$ spin Calogero model}
In this section we will introduce the $osp(m|n)$ spin Calogero models.
Let $V$ be an $m+n$ dimesional $ \mathbb{Z}_2$ graded vector space
and $\{ v^{a}, a=1,\dots,m+n \}$ be a homogeneous basis whose grading
is as same as before:
\begin{displaymath}
 [a] = \left\{
         \begin{array}{ll}
           0, & a=1, \dots ,m \\
           1, & a=m+1, \dots , m+n.
        \end{array}
        \right. 
\end{displaymath}
We consider $L$ copies of the generators of $gl(m|n)$  
$e_{j}^{ab} \,(j=1, \dots , L)$ that act on the
$j$-th space of the tensor product of graded vector spaces
$V_{1} \otimes \cdots \otimes V_{L}$ where the subscript $j$ corresponds
to the space $V_{j} \simeq V$ in the tensor product. With the relation
\begin{equation}
 (e_{j}^{ab} \otimes e_{k}^{cd}) v_{j}^{p} \otimes v_{k}^{q}=
  (-1)^{([c]+[d])[p]}e_{j}^{ab}v_{j}^{p} \otimes e_{k}^{cd} v_{k}^{q},
\end{equation}
one can show that the permutation operator $P_{jk}$ defined as
\begin{equation}
 P_{jk}=\sum_{a,b=1}^{m+n}(-1)^{[b]}e_{j}^{ab} \otimes e_{k}^{ba}
\end{equation}
exchanges the spin state of the $j$-th particle $v_{j}^{a}$ and 
the $k$-th particle $v_{k}^{b}$ . 
Furthermore we introduce an operator $Q_{jk}$ as follows:
\begin{equation}
 Q_{jk}=\sum_{a,b=1}^{m+n} \xi_{a} \xi_{b} (-1)^{[a][b]} 
 e_{j}^{ab} \otimes e_{k}^{{\bar a}{\bar b}}.
\end{equation}
The actions of these operators on $v_{j}^{a} \otimes v_{k}^{b}$ are explicitly 
written as
\begin{eqnarray}
 P_{jk} v_{j}^{a} \otimes v_{k}^{b} &=& (-1)^{[a][b]}v_{j}^{b} \otimes v_{k}^{a}, \\
 Q_{jk} v_{j}^{a} \otimes v_{k}^{b} &=& 
 \delta_{a {\bar b}} \sum_{c=1}^{m+n} \xi_{c} \xi_{{\bar a}} 
  v_{j}^{c} \otimes v_{k}^{{\bar c}}.
 \end{eqnarray}
They satisfy the usual properties $P_{jk}=P_{kj}$ and $Q_{jk}=Q_{kj}$.
Now we consider the following Hamiltonian
\begin{equation}
 H^{(m|n)} = - \sum_{j=1}^{L} \frac{\partial^2}{\partial x_{j}^{2}}
  + 2 \lambda \sum_{j < k} 
 \frac{\left(\lambda -\left(P_{jk}-Q_{jk}\right) \right)}{(x_{j}-x_{k})^2}.
\label{osphamiltonian}
\end{equation}
The operator $P_{jk}-Q_{jk}$ is the exchange operator interchanging the "spins" of
$j$-th and $k$-the lattice site.
Note that we can write the new interactions in terms of $osp(m|n)$ generators as
follows
\begin{equation}
 P_{jk}-Q_{jk} =-\frac{1}{2} \sum_{a,b=1}^{m+n} \xi_{a} \xi_{b}(-1)^{[a][b]}
 \sigma_{j}^{ab} \sigma_{k}^{{\bar a}{\bar b}}.
\end{equation}
In this sense we call the models described by the Hamiltonian (\ref{osphamiltonian}) 
$osp(m|n)$ spin Calogero models.

\section{Symmetry of $osp(m|n)$ spin Calogero models}
In this section we will obtain the symmetry of the $osp(m|n)$ spin Calogero models.
For this purpose, we introduce the following two operators
\begin{eqnarray}
 J_{0}^{ab} &=& \sum_{j=1}^{L} \sigma_{j}^{ab}, \\
 J_{1}^{ab} &=& \sum_{j=1}^{L} \sigma_{j}^{ab} \frac{\partial}{\partial x_{j}}
  - \lambda \sum_{j \ne k} (\sigma_{j} \sigma_{k})^{ab} \frac{1}{x_{j}-x_{k}}.
\end{eqnarray}
Here we have used the notations,
\begin{equation}
 (\sigma_{j} \sigma_{k})^{ab}=\sum_{c=1}^{m+n} \xi_{c} \sigma_{j}^{ac} 
   \sigma_{k}^{{\bar c} b}.
\end{equation}
By simple calculation we collect various useful formulas: For $j \ne k \ne l \ne m$,
\begin{eqnarray}
 \left[ P_{jk}-Q_{jk}, \sigma_{l}^{ab} \right] &=& 0, \\
 \left[ P_{jk}-Q_{jk}, \sigma_{k}^{ab} \right] &=& -(\sigma_{j}\sigma_{k})^{ab}
  +(-1)^{[a][b]}(\sigma_{j}\sigma_{k})^{ba} \\
 \left[ P_{jk}-Q_{jk},(\sigma_{l}\sigma_{m})^{ab} \right] &=& 0, \\
 \left[ P_{jk}-Q_{jk},(\sigma_{j}\sigma_{l})^{ab} \right] &=& 
  -(\sigma_{j}\sigma_{k}\sigma_{l})^{ab} + (\sigma_{k}\sigma_{j}\sigma_{l})^{ab}, \\
 \left[ P_{jk}-Q_{jk},(\sigma_{j}\sigma_{k})^{ab} \right] &=&
 -(\sigma_{j}\sigma_{k}\sigma_{k})^{ab}+(\sigma_{k}\sigma_{j}\sigma_{k})^{ab} \nonumber \\
 && \quad +(\sigma_{j}\sigma_{j}\sigma_{k})^{ab}-(\sigma_{j}\sigma_{k}\sigma_{j})^{ab},
\end{eqnarray}
where we have defined
\begin{equation}
 (\sigma_{j} \sigma_{k} \sigma_{l})^{ab}=\sum_{p,q=1}^{m+n} \xi_{p}\xi_{q} 
  \sigma_{j}^{ap} \sigma_{k}^{{\bar p}q} \sigma_{l}^{{\bar q}b}.
\end{equation}
In addition the following formulas are also useful. For $j \ne k \ne l$,
\begin{eqnarray}
 (\sigma_{k}\sigma_{j})^{ba} &=& (-1)^{[a][b]}(\sigma_{j}\sigma_{k})^{ab}, \\
 (\sigma_{j}\sigma_{k}\sigma_{l})^{ba} &=& 
   -(-1)^{[a][b]}(\sigma_{l}\sigma_{k}\sigma_{j})^{ab},\\
 (\sigma_{k}\sigma_{k}\sigma_{j})^{ba} &=& 
   (-1)^{[a][b]}(\sigma_{j}\sigma_{k}\sigma_{k})^{ab}-
   (m-n-2)(-1)^{[a][b]}(\sigma_{j}\sigma_{k})^{ab}, \\
 (\sigma_{k}\sigma_{j}\sigma_{k})^{ba} &=&
   -(-1)^{[a][b]}(\sigma_{k}\sigma_{j}\sigma_{k})^{ab}  \nonumber \\
   && \quad -g_{ba} \sum_{p,q=1}^{m+n}\xi_{p}\xi_{q}(-1)^{([a]+[q])([b]+[q])}
             \sigma_{k}^{{\bar q}p}\sigma_{j}^{{\bar p}q}.
\end{eqnarray}
Then the followings are results of this section.

\newtheorem{prop}{Proposition}
\begin{prop}
The generators $J_{0}^{ab}$ and $J_{1}^{ab}$ satisfy the following 
relations
\begin{eqnarray}
 [ J_{0}^{ab}, J_{0}^{cd} ] &=& g_{cb} J_{0}^{ad}
  -(-1)^{([a]+[b])([c]+[d])}g_{ad}J_{0}^{cb}  \nonumber \\
   &&-(-1)^{[c][d]}(g_{db} J_{0}^{ac} 
   -(-1)^{([a]+[b])([c]+[d])}g_{ac}J_{0}^{db} ),
 \label{loop1}
\end{eqnarray}
\begin{eqnarray}
 [ J_{0}^{ab}, J_{1}^{cd} ] &=& g_{cb} J_{1}^{ad}
  -(-1)^{([a]+[b])([c]+[d])}g_{ad}J_{1}^{cb}  \nonumber \\
   &&-(-1)^{[c][d]}(g_{db} J_{1}^{ac} 
   -(-1)^{([a]+[b])([c]+[d])}g_{ac}J_{1}^{db} ), 
  \label{loop2}
\end{eqnarray}
\begin{eqnarray}
 &&(-1)^{([a]+[b])([c]+[d])}\left[J_{1}^{cd},
    [J_{0}^{ab}, J_{1}^{ef}] \right] \nonumber \\
 && \quad +\left[ [J_{0}^{ab},J_{1}^{cd} ],J_{1}^{ef} \right]
      -\left[J_{1}^{ab},[J_{0}^{cd},J_{1}^{ef}] \right]=0,  
    \label{loop3}   
\end{eqnarray}
for the following particular value of the coupling constant
\begin{equation}
 \lambda =\frac{2}{m-n-4}.
\label{coupling}
\end{equation}
\end{prop}

{\it Proof.} \enskip
The first and the second relations can be shown by straightforward
calculations.
In order to prove the third relation, 
we compute $\left[J_{1}^{ab}, J_{1}^{cd} \right]$.
Then we obtain 
that if the coupling constant $\lambda$ equals to 
(\ref{coupling}), then 
\begin{eqnarray}
 [ J_{1}^{ab}, J_{1}^{cd} ] &=& g_{cb} J_{2}^{ad}
  -(-1)^{([a]+[b])([c]+[d])}g_{ad}J_{2}^{cb}  \nonumber \\
   &&-(-1)^{[c][d]}(g_{db} J_{2}^{ac} 
   -(-1)^{([a]+[b])([c]+[d])}g_{ac}J_{2}^{db} ),
\end{eqnarray}
where we define 
\begin{eqnarray}
 && J_{2}^{ab} = \nonumber \\
 && \sum_{j=1}^{L}\sigma_{j}^{ab} \frac{\partial^2}{\partial x_{j}^2}-
 \lambda \sum_{j \ne k} (\sigma_{j} \sigma_{k})^{ab} 
 \frac{1}{x_{j}-x_{k}} \left( \frac{\partial}{\partial x_{j}}+
 \frac{\partial}{\partial x_{k}} \right) \nonumber \\
 &&  + \lambda \sum_{j \ne k} \left\{ -\lambda \sigma_{j}^{ab}
   -\lambda \sigma_{k}^{ab}
 +(\sigma_{j} \sigma_{k} \sigma_{j})^{ab}- (-1)^{[a][b]}
 (\sigma_{k} \sigma_{j} \sigma_{k})^{ab}\right\}\frac{1}{(x_{j}-x_{k})^2} \nonumber \\
 &&  + \lambda^2 \sum_{j \ne k \ne l} (\sigma_{j} \sigma_{k} \sigma_{l})^{ab}
 \frac{1}{x_{j}-x_{k}} \frac{1}{x_{k}-x_{l}}.
\end{eqnarray} 
Consequently the super Jacobi identity (\ref{superJ}) assures 
the third relation of the proposition.
\hfill  $\Box$

The equation (\ref{loop3}) is called Serre relation for the loop algebra.
Thanks to (\ref{loop3}) we can define the higher level generators 
$J_{2}^{ab}, \, J_{3}^{ab}, \, \cdots$ recursively:
\begin{equation}
 J_{\nu}^{ab}=\frac{1}{|f_{cd, ef, ab}f_{ef, cd, ab}|}f_{cd, ef,ab} 
 [J_{1}^{cd}, J_{\nu-1}^{ef}],
\end{equation}
where $f_{ab, cd, ef}$ are the structure constants of $osp(m|n)$, namely
\begin{equation}
 [\sigma^{ab}, \sigma^{cd}]=f_{ab,cd,ef} \sigma^{ef}.
\end{equation}
These relations (\ref{loop1})-(\ref{loop3}) imply the generators 
$J_{\nu}^{ab} \, (\nu \ge 0)$ form the half loop algebra
associated to the $osp(m|n)$,
\begin{eqnarray}
 \left[ J_{\mu}^{ab}, J_{\nu}^{cd} \right]&=&
  g_{cb} J_{\mu+\nu}^{ad}
  -(-1)^{([a]+[b])([c]+[d])}g_{ad}J_{\mu+\nu}^{cb} \nonumber \\
  && -(-1)^{[c][d]}(g_{db} J_{\mu+\nu}^{ac} 
   -(-1)^{([a]+[b])([c]+[d])}g_{ac}J_{\mu+\nu}^{db} ).
\end{eqnarray}
The next proposition shows that the generators of the $osp(m|n)$ half loop
algebra $J_{\nu}^{ab}$ are conserved operators for the $osp(m|n)$ spin
Calogero model.
\begin{prop}
The operators $J_{0}^{ab}$ and $J_{1}^{ab}$ commute with the Hamiltonian of $osp(m|n)$ spin
Calogero model $H^{(m|n)}$ :
\begin{equation}
 \left[ H^{(m|n)}, J_{0}^{ab} \right] =0, \quad
 \left[ H^{(m|n)}, J_{1}^{ab} \right] =0, 
\end{equation}
for the coupling constant $\lambda$ equals to (\ref{coupling}).
\end{prop}
Therefore we conclude that the symmetry algebra of the model described by 
the Hamiltonian (\ref{osphamiltonian}) is the half-loop algebra associated
to $osp(m|n)$ if and only if the coupling constant $\lambda$ equals to 
$\frac{2}{m-n-4}$.

\section{$osp(m|n)$ spin Sutherland models}

We naturally expect that $osp(m|n)$ spin Sutherland model, whose Hamiltonian
given by
\begin{equation}
 H^{(m|n)}_{Suth} = - \sum_{j=1}^{L} \frac{\partial^2}{\partial {\xi}_{j}^{2}}
  + \frac{\lambda}{2} \, \sum_{j < k} 
 \frac{\left(\lambda -\left(P_{jk}-Q_{jk}\right) \right)}
        {\sin^2\left[({\xi}_{j}-{\xi}_{k})/2 \right]},
\label{ospsutherland}
\end{equation}
have the symmetry of Yangian $Y(osp(m|n))$. 
In order to see this we first rewrite the Hamiltonian (\ref{ospsutherland})
in terms of the variables $x_{j}={\rm exp}(\sqrt{-1} \, \xi_{j})$.
Then we have
\begin{equation}
 \widehat{H}^{(m|n)}_{Suth} = \sum_{j=1}^{L} 
 \left(x_{j} \frac{\partial}{\partial x_{j}} \right)^{2}
  - 2 \lambda \sum_{j < k} 
  \left(\lambda -\left(P_{jk}-Q_{jk}\right) \right) 
  \frac{x_{j}x_{k}}{(x_{j}-x_{k})^{2}}.
\label{cvospsutherland}
\end{equation}

Next we introduce a new set of operators as follows:
\begin{eqnarray}
 K_{0}^{ab} &=& \sum_{j=1}^{L} \sigma_{j}^{ab}, \\
 K_{1}^{ab} &=& \sum_{j=1}^{L} \sigma_{j}^{ab} 
                \left(x_{j}\frac{\partial}{\partial x_{j}}\right)
  - \frac{\lambda}{2} \sum_{j \ne k} (\sigma_{j} \sigma_{k})^{ab} 
     \frac{x_{j}+x_{k}}{x_{j}-x_{k}}.
\end{eqnarray}
Then we obtain the following results for the $osp(m|n)$ spin Sutherland models.

\begin{prop}
The generators $K_{0}^{ab}$ and $K_{1}^{ab}$ satisfy the following commutation
relations when the coupling constant $\lambda$ equals to (\ref{coupling}).
\begin{eqnarray}
 [ K_{0}^{ab}, K_{0}^{cd} ] &=& g_{cb} K_{0}^{ad}
  -(-1)^{([a]+[b])([c]+[d])}g_{ad}K_{0}^{cb}  \nonumber \\
   &&-(-1)^{[c][d]}(g_{db} K_{0}^{ac} 
   -(-1)^{([a]+[b])([c]+[d])}g_{ac}K_{0}^{db} ),
 \label{yangian1}
\end{eqnarray}
\begin{eqnarray}
 [ K_{0}^{ab}, K_{1}^{cd} ] &=& g_{cb} K_{1}^{ad}
  -(-1)^{([a]+[b])([c]+[d])}g_{ad}K_{1}^{cb}  \nonumber \\
   &&-(-1)^{[c][d]}(g_{db} K_{1}^{ac} 
   -(-1)^{([a]+[b])([c]+[d])}g_{ac}K_{1}^{db} ), 
  \label{yangian2}
\end{eqnarray}
\begin{eqnarray}
 &&(-1)^{([a]+[b])([c]+[d])}\left[K_{1}^{cd},
    [K_{0}^{ab}, K_{1}^{ef}] \right] \nonumber \\
 && \quad\quad +\left[ [K_{0}^{ab},K_{1}^{cd} ],K_{1}^{ef} \right]
      -\left[K_{1}^{ab},[K_{0}^{cd},K_{1}^{ef}] \right] \nonumber \\
  &&=\frac{\lambda^2}{4}\Big\{(-1)^{([a]+[b])([c]+[d])}
      (K_{0}K_{0}K_{0})^{[cd,[ab,ef]]}  \label{yangian3} \\
      && \quad\quad +(K_{0}K_{0}K_{0})^{[[ab,cd],ef]}
      -(K_{0}K_{0}K_{0})^{[ab,[cd,ef]]} \Big\}. \nonumber
\end{eqnarray}
Here we use the following notations.
\begin{eqnarray}
 (K_{0}K_{0}K_{0})^{[ab,[cd,ef]]}&=&g_{ed}(K_{0}K_{0}K_{0})^{ab,cf} \nonumber \\
&&-(-1)^{([c]+[d])([e]+[f])}g_{cf}(K_{0}K_{0}K_{0})^{ab,ed}\\
&&-(-1)^{[e][f]}g_{fd}(K_{0}K_{0}K_{0})^{ab,ce} \nonumber \\
&&+(-1)^{[e][f]+([c]+[d])([e]+[f])}g_{ce}(K_{0}K_{0}K_{0})^{ab,fd}, \nonumber
\end{eqnarray}
and
\begin{eqnarray}
 (K_{0}K_{0}K_{0})^{ab,cd} &=& (-1)^{[b][c]}(K_{0}K_{0})^{ac}K_{0}^{bd} \nonumber \\
 &&+(-1)^{[b][c]+[a][b]+[a][c]}K_{0}^{cb}(K_{0}K_{0})^{ad} \\
 && -(-1)^{[b][c]+[a][b]+[a][c]}(K_{0}K_{0})^{cb}K_{0}^{ad} \nonumber \\
 &&+(-1)^{[b][c]+[a][c]+[b][d]}K_{0}^{ca}(K_{0}K_{0})^{db}. \nonumber
\end{eqnarray}
\end{prop}

The relations (\ref{yangian1})-(\ref{yangian3}) are 
the defining relations of the super Yangian $Y(osp(m|n))$. The equation (\ref{yangian3}) 
is called the deformed Serre relation for the super Yangian. Note that it reduces to
the Serre relation (\ref{loop3}) for the loop algebra in the limit of $\lambda \to 0$.

One then directly show the next proposition.

\begin{prop}
The operators $K_{0}^{ab}$ and $K_{1}^{ab}$ are conserved operators for the
$osp(m|n)$ spin Sutherland model, that is, they commute with 
the Hamiltonian $\widehat{H}^{(m|n)}_{Suth}$ :
\begin{equation}
 \left[ \widehat{H}^{(m|n)}_{Suth}, K_{0}^{ab} \right] =0, \quad
 \left[ \widehat{H}^{(m|n)}_{Suth}, K_{1}^{ab} \right] =0, 
\end{equation}
if the coupling constant $\lambda$ equals to (\ref{coupling}).
\end{prop}

It follows from these that the $osp(m|n)$ spin Sutherland models have
the super Yangian symmetry $Y(osp(m|n))$ when the coupling constant 
$\lambda$ equals to $\frac{2}{m-n-4}$.

\bigskip

\noindent{\bf Acknowledgement}
 
\medskip

\noindent I am grateful to Professors Hidetoshi Awata, Hiroyuki Yamane and Ken Ito 
for their kind encouragements and helpful comments.

\end{document}